\documentclass[aip,reprint,graphicx]{revtex4-1}

\usepackage{amsmath,amssymb}
\usepackage{graphicx}
\usepackage{dcolumn}
\usepackage{bm}
\usepackage{booktabs,ragged2e}
\usepackage[mathscr]{euscript}

\usepackage[utf8]{inputenc}
\usepackage[T1]{fontenc}
\usepackage{mathptmx}
\usepackage{etoolbox}

\draft 

\begin{document}


\title{A quantitative analysis of stochastic resonance in ferroelectrics} 



\author{Madhav Ramesh}
  \email{mr974@cornell.edu}
 \altaffiliation[Currently at ]{School of Electrical and Computer Engineering, Cornell University, Ithaca, New York 14853, USA}
 
\author{Amit Verma}
 \email{amitkver@iitk.ac.in}
\altaffiliation{Department 
of Electrical Engineering, Indian Institute of Technology Kanpur, Kanpur 208016, Uttar Pradesh, India
}

\author{Arvind Ajoy}
 \email{arvindajoy@iitpkd.ac.in}
\altaffiliation{Department 
of Electrical Engineering, Indian Institute of Technology Palakkad, Palakkad 678623, Kerala, India
}

%


\date{\today}

\begin{abstract}
A simulation-based study of stochastic resonance (SR) in a ferroelectric capacitor is presented.  The SR phenomenon involves the detection of weak signals by adding an optimal amount noise to a non-linear system.  This is linked with Kramers' escape problem, which deals with the escape of a particle undergoing Brownian motion over an energy barrier.  The position of the particle is analogous to the polarisation dynamics of a ferroelectric. Within this framework, we numerically investigate SR in single domain ferroelectrics using the Landau-Ginzburg-Devonshire (LGD) theory.  In addition, we use a model for multidomain ferroelectrics to demonstrate feasibility in real world applications. Our results show that SR in ferroelectrics is promising for the purpose of weak signal detection,  given that these materials are widely used for various applications in the semiconductor industry. 
\end{abstract}


\maketitle 

\section{Introduction}
 Stochastic Resonance (SR) is a counter-intuitive phenomenon wherein weak periodic signals given as input to systems which have an inherent threshold or energetic bistability,  may be detected with the addition of an optimal amount of noise.  In these systems, Signal-to-Noise Ratio (SNR) increases with an  increase in noise intensity, within a certain range of noise intensities.  It was first studied by Benzi et al.~\cite{benzi_tellus_1982} using a climate change model to interpret the atmospheric temperature variations over large time periods.  Subsequently, it was discovered in the nervous systems of crayfish and paddlefish as an evolutionary mechanism, resulting in its importance in neurobiological research~\cite{douglass_nature_1993, russell_nature_1999}. Recently, several engineering applications which use this phenomenon have also emerged, such as a visual aid~\cite{itzcovich_scirep_2017}, a low power photodetector~\cite{dodda_natcom_2020}, an Insulator-Metal Transition (IMT) based system for ultra-low power auditory processing~\cite{bhar_scirep_2020}. 
\par 
Fundamentally, SR can be traced back to Kramers' escape problem, which deals with the dynamics of the escape of a Brownian particle from a potential well, over an energy barrier \cite{kramers_physica_1940,hanggi_revmodphys_1990,hanggi_jstatphys_1986}. Brownian motion \cite{einstein_cc_1956} is the stochastic process which gives rise to  thermal noise in electronic systems, also known as Johnson-Nyquist noise \cite{johnson_physrev_1928,nyquist_physrev_1928}. It is white, implying that its power spectral density is flat as a function of frequency. For the escape problem, the case where the particle is in a strongly damped environment subject to thermal noise is of significant interest. The equations (provided in the next section of this paper) governing the dynamics of the unknown stochastic quantity (in this case, position of the Brownian particle), closely resemble the polarisation dynamics of a ferroelectric in accordance with the Landau-Ginzburg-Devonshire (LGD) theory \cite{landau_eksp_1937,ginzburg_eksp_1945,devonshire_philmag_1949}.  An analogue of the classic escape problem under these conditions, therefore, is one that deals with electric polarisation as the unknown stochastic quantity, rather than position. Since the defining property of a ferroelectric is spontaneous polarisation, it is possible to map Kramers' problem (and therefore SR) to ferroelectrics to understand its utility for engineering applications.  \par
Few papers \cite{arai_iscas_2021,hakamata_JAppPhys_2010} have presented device applications for weak electronic signal detection. However, none of them incorporate ferroelectrics, which have an inherent bistable well. It is to be noted that SR has been experimentally reported in  ferroelectric Triglycine Sulphate (TGS) \cite{drozhdin_phdthesis} -- however, this work does not present a quantitative comparison of the experimental data with theoretical predictions. Additionally, it has not been motivated as an application of Kramers' problem  to detect weak periodic signals. 

In order to make use of this analogy, a fundamental assumption is that all the domains of the ferroelectric behave homogeneously in the presence of an applied electric field, i.e., we approximate the material as a single domain. 
Under this assumption, the polarisation switching dynamics is described by the Time Dependent Ginzburg-Landau (TDGL) equation. This is commonly referred to as homogeneous or intrinsic switching, as against extrinsic switching which is largely governed by nucleation and domain growth \cite{vizdrik_prb_2003}.  Although the extrinsic mechanism is more common, homogeneous switching has been experimentally observed in certain ferroelectrics at scaled thicknesses, such as P(VDF-TrFE) \cite{gaynutdinov_JPhys_condmat_2011},  PVDF \cite{tian_PRB_2015} and PbTiO\textsubscript{3} \cite{highland_physrevlett_2010}.  

In this work, we consider  an HfO\textsubscript{2}-based thin-film, Hafnium Zirconium Oxide (Hf\textsubscript{1-x}Zr\textsubscript{x}O\textsubscript{2} or HZO) as the ferroelectric. Applications based on HZO could potentially have commercial significance, owing to this material's compatibility with the existing CMOS processes  \cite{kim_jom_2019, si_nature_nano_2018,muller_apl_2011,Dunkel_iedm_2017}.  It is still an ongoing debate as to whether switching in HZO is truly intrinsic or whether the domain growth contributes to the behaviour \cite{hoffmann_AFM_2021}.  However, there have been recent reports of intrinsic switching in polycrystalline Si:HfO\textsubscript{2} \cite{gastaldi_apl_2021} and HZO \cite{stolichnov_apl_2020}, suggesting that such a mechanism may be possible.  More recently, it was shown that in Y:HfO\textsubscript{2} capacitors (both epitaxial and polycrystalline), the inhomogeneous switching mechanisms (Kolmogorov Avrami Ishibashi, KAI   \cite{kolmogorov_bullacadsci_1937,avrami_JChemPhys_1939,ishibashi_Jphyssocjap_1971} and Nucleation Limited Switching, NLS \cite{tagantsev_PRB_2002}) converge for high electric fields, making switching resemble the homogeneous process \cite{buragohain_AFM_2021}.  
 
Under the assumption of homogeneous switching, it is possible to use the LGD theory to model the ferroelectric as a single domain and understand its relation to the escape problem.   Etesami et al. \cite{etesami_phys_rev_b_2016} demonstrated the impact of thermal fluctuations using single and multidomain ferroelectric models. Here, we present the theory and simulations for SR instead.  Although we are using HZO parameters for our simulations,  our analysis is transferrable to any material system with bistable behaviour.  In addition, we display simulation results using a model to account for the multidomain/inhomogeneous nature of ferroelectrics. To shed further light on the feasibility of the single domain assumption, we also demonstrate a match between the results from both models. This opens up the possibility of realising potentially useful engineering applications based on ferroelectric materials for the purpose of weak signal detection.
 
\par

In this paper, we present an analysis of SR in a ferroelectric capacitor. We hope to convey its relevance to the devices community through theory and numerical simulations. This would involve an initial introduction to Kramers' escape problem and the underlying mathematics, followed by simulation results for both single and multidomain ferroelectrics.

\section{Overview of Kramers' escape problem}
\begin{figure}[t]
\centering
\includegraphics[scale=0.8]{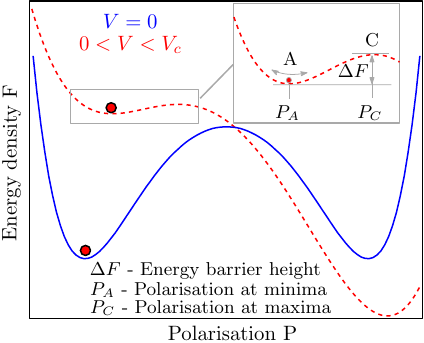}
\caption{Kramers' problem for a double well potential landscape in a ferroelectric. The blue(solid) line depicts the double well potential when a voltage $V=0$ is applied across the ferroelectric, while the red(dashed) line depicts the situation under an applied voltage $0<V<V_c$. $V_c$ is the coercive voltage where the potential barrier disappears.  The state of polarisation denoted by the red disc is analogous to a particle trapped in a potential well. Assuming the polarisation is at point A when $V < V_c$,  it can still escape the well, over  point C, when subject to thermal fluctuations. Kramers time quantifies the average lifetime of system in state A.}
\label{fig:kramersproblem}
\end{figure} 
In 1940, Kramers provided the mathematical framework to describe the escape of a particle over a potential barrier, in the context of dynamics of chemical reactions \cite{kramers_physica_1940}. The dynamics of a classical particle of mass $M$ in a potential $U(x)$, with damping $\gamma$,  can be summarized as
\begin{equation}\label{eq:krampart}
M\ddot{x} = -\frac{\partial U(x)}{\partial x} - \gamma \dot{x} + \eta(t)
\end{equation}
where $x$ is the position,  and $\eta(t)$ is a fluctuating force described by Gaussian-white noise. This is similar to the equation for a spring-mass damper,  with an additional fluctuating force. In the strongly damped limit, this equation reduces to Langevin's equation \cite{langevin_CRPhys_1908},
\begin{equation}\label{eq:kramdamped}
 \gamma \dot{x} = -\frac{\partial U(x)}{\partial x} + \eta(t)
\end{equation}
since the inertial term may be neglected. This particular equation maps to the equation that governs the spontaneous polarisation dynamics of a ferroelectric. 

Consider the double well potential in Figure \ref{fig:kramersproblem}. Given this energy landscape for the ferroelectric, one may observe this is similar to the setup for Kramers' problem with electric polarisation being the stochastic variable (instead of position $x$). Let the state of the system initially be at point A. We are then interested in the average time and the probability for which the system reaches C and ultimately crosses the barrier. Note that the switching dynamics of the system can be altered by applying a voltage, which changes the relative depths of the two potential wells. 

Since thermal noise has already been studied in a ferroelectric \cite{etesami_phys_rev_b_2016}, we use the same model to define the problem at hand. The free energy density $F$ (in $J/m^3$) of the ferroelectric is related to the polarisation $P$ (under the assumption of homogeneous switching) and is given by
\begin{equation}\label{eq:energy}
F = \alpha P^2 + \beta P^4 -PE,
\end{equation}
where $E$ is the electric field and $\alpha$ and $\beta$ are the Landau coefficients. 
Then, the TDGL equation is
\begin{equation}\label{eq:langvn}
\rho \frac{\partial P}{\partial t} = -\frac{\partial F}{\partial P} + \xi (t),
\end{equation}
where  $\xi(t)$ is the noise (a fluctuating term) and $\rho$ is the resistivity (a dissipative term). Note that this resembles the Langevin equation. Further note that this equation represents a class of Stochastic Differential Equations (SDEs) which model a system that has a fluctuating term  along with a dissipative term. Assuming Gaussian white noise, the auto-correlation of the noise is given by
\begin{equation}\label{eq:flucdis1} 
\langle \xi (t)\xi (t')\rangle = \frac{2k_BT\rho}{A_F t_F} \delta (t-t'),
\end{equation}
from the Fluctuation dissipation relation, which relates the fluctuating and dissipative forces in the system. Here, $k_B$ is Boltzmann's constant and $T$ is the temperature of the system. Further, $t_F$ and $A_F$ represent the thickness and area respectively of the ferroelectric under consideration. These terms play the important role of determining how significant internal noise is in the system. Highly scaled ferroelectrics (i.e. low $t_F$) exhibit more thermal noise, as is evident from the autocorrelation expression in eq. (\ref{eq:flucdis1}).  \par


In order to define a scale for the noise in the problem, we consider the Fokker-Planck equation or FPE \cite{risken_springer_1996} to describe the evolution of the probability density function $w(P,t)$.  The FPE for the situation under consideration is given as
\begin{equation}\label{eq:fokkpl}
\frac{\partial}{\partial t} w(P, t)=\frac{1}{\rho}\frac{\partial}{\partial P}\left[w \frac{\partial F}{\partial P}+D \frac{\partial w}{\partial P}\right]
\end{equation}
where $D$ is the diffusion constant. We exploit the fact that at equilibrium, $w(P,t)$ is a Boltzmann distribution. Setting the probability current $\frac{\partial}{\partial t} w(P, t)=0$ in eq. (\ref{eq:fokkpl}) at equilibrium yields  $D = \frac{k_BT}{t_F A_F}$. Note that the $t_F A_F$ term is the volume and it emerges because the energy density F is in units of $J/m^3$, or energy per unit volume. The noise $\xi(t)$ can then be written in terms of a Brownian motion or Wiener Process $W(t)$ \cite{levy_springe_2020}
\begin{equation}\label{eq:Wiener}
\xi(t) = \sqrt{2  \rho D} \; \frac{dW(t)}{dt}.
\end{equation}

So far, in eq. (\ref{eq:flucdis1}, \ref{eq:Wiener}), there is no notion of external noise being added to the system. We now include an external noise voltage (with root mean squared value $V_{noise}$) having a power spectral density that is  flat over a bandwidth $\Delta f$. This external noise is uncorrelated to the internal noise in the system. Hence, eq. (\ref{eq:Wiener}) can be modified as 
\begin{equation}\label{eq:WienerExternal}
\xi(t) = \left( \sqrt{2  \rho D_{int}} + \sqrt{2  \rho D_{ext}}\right)\; \frac{dW(t)}{dt}
\end{equation}
where
\begin{align}
\label{eq:D}
D_{int} &=\frac{k_BT}{t_F A_F} \\
D_{ext} &= \frac{V_{noise}^2}{2R_F} \frac{1}{\Delta f} \frac{1}{t_F A_F}, \text{with } R_F = \frac{\rho t_F}{A_F}
\end{align}
represent the effect of internal and external noise respectively.  This definition is  similar to the description of a magnetic tunnel junction with stochastic input, as discussed in \cite{liyanagedera_PRA_2017}.
\par

We reiterate that the  SDE eq. (\ref{eq:langvn}) is analogous to the classic escape problem under strongly-damped conditions. Following Metzler and Klafter \cite{metzler_chem_phys_lett_2000}, we can determine the rate of escape of the state of polarisation over the energy barrier. This is known as Kramers rate $r_K$ where
\begin{equation}\label{eq:krrate}
r_K = \frac{1}{t_K} = \frac{\sqrt{|F''(P_{A})F''(P_{C})|}}{2\pi\rho}\exp\left({-\dfrac{\Delta F}{D_{ext}}}\right)
\end{equation}
under the assumption that $D_{ext}\gg D_{int}$. Refer to Figure \ref{fig:kramersproblem} for the notations. $F''$ denotes the second derivative of the free energy density. The reciprocal of $r_K$ is $t_K$, which is referred to as Kramers time. This particular metric can be interpreted as the average time spent in the well around point A, before a transition is made over point C at the top of the barrier. Note finally that the ratio $D_{ext}/ \Delta F$ naturally provides a scale for the noise, since this ratio determines Kramers time.

\section{Simulation results and discussion}
The ferroelectric parameters and thickness values for the simulations have been determined from the remnant polarisation ($P_r = 17.76$ $\mu$C/cm$^2$) and coercive electric field ($E_c=104$ MV/m, corresponding to $V_c = 1.04$ V) reported for  $t_F = 10$ nm thick Hf\textsubscript{0.5}Zr\textsubscript{0.5}O\textsubscript{2} in \cite{gaddam_IEEE_trans_ED_2020}. The $\alpha$ and $\beta$ values were obtained using the following expressions \cite{lin_TED_2016}
\begin{equation}\label{eq:alphabeta}
\alpha = \frac{-3\sqrt{3}E_c}{4P_r} \text{ and } \beta = \frac{3\sqrt{3}E_c}{8P_r^3}
\end{equation}  
The value of resistivity $\rho$ is taken from experimental data\cite{si_nature_nano_2018}, while the area has been chosen arbitrarily but in the range of existing values in literature. All relevant parameters have been listed in Table \ref{tab:ferroparam}. 
\subsection{Simulation methodology}
We solve eq. (\ref{eq:langvn}) using the Euler-Maruyama method, similar to the  Euler method that is used to solve ordinary differential equations \cite{platen_numerica_1999}.  In discrete form, with $D_{ext}\gg D_{int}$, we have 
\begin{align}\label{eq:algo}
P[i] &= P[i-1] - \frac{\Delta t}{\rho} \cdot \frac{ dF}{dP}[i] + \sqrt{\frac{2 D_{ext}}{\rho}} \cdot \Delta W[i]
\end{align}
where $[i]$ represents the $i^{th}$ time-step.  
The term $\Delta W[i]$ is obtained by extracting numbers from a normal distribution with mean 0 and variance $\Delta t$. For our simulations, we selected a time step $\Delta t = $ 1 ns. Note that the time step for Brownian dynamics simulations must be as small as possible, so as to prevent the possibility of divergence for high noise intensities. 

\begin{table}[t]
    \caption{Parameters of HZO capacitors used in this work}
    \label{tab:ferroparam}    
    \centering
    \begin{ruledtabular}
    \begin{tabular}{cc}
        \textbf{Parameter} & \textbf{Value} \\
        \hline
        Thickness $t_F$ ($nm$) & $10$ \\
         $\alpha$ ($m F^{-1}$) &  $-7.603\times10^8$ \\
         $\beta$ ($m^5 F^{-1} C^{-2}$) &  $1.204\times10^{10}$ \\
         Resistivity $\rho$ ($\Omega - m$) & $30$ \\
         Temperature T ($K$) & $300$ \\
         Area $A_F$ ($\mu m^2$) & $1$ \\
    \end{tabular}
    \end{ruledtabular}
\end{table}
\begin{figure}[b]
\centering
\includegraphics[scale=0.8]{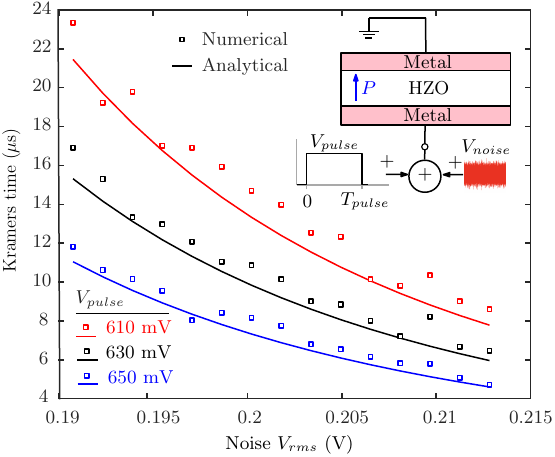}
\caption{Verification of Kramers time through numerical simulations. Our results match the analytical predictions, based on eq. (\ref{eq:krrate}), very well.}
\label{fig:kramerstimeverification1}
\end{figure} 
Figure \ref{fig:kramerstimeverification1} presents the results of our simulations for Kramers time $t_K$, compared with the analytical results predicted by eq. (\ref{eq:krrate}).  The polarization is initially assumed to be in the left well. A positive voltage pulse with amplitude $V_{pulse}$ is applied such that the double well is asymmetric. A significantly large pulse width $T_{pulse}=90\mu s$ has been taken (about 4$\times$ larger than the highest value of Kramers time) such that a switching event is almost certainly guaranteed.  The asymmetry has  been chosen such that the average time spent in the left well is finite, whereas the right well yields a very large time approaching infinity. An ensemble of 300 systems have been considered (for each of the $V_{rms}$ and $V_{pulse}$ values), with the average time spent in the well before a switching event occurring being recorded.  Based on our simulation setup, switching can never occur from the right well to the left well, ensuring that the numerically calculated average does not include any unwanted reverse switching events. The good match between the analytical and numerical results verifies the working of our numerical solver. This solver will be used to investigate SR through single ferroelectric domain switching. 
\subsubsection{Linear response theory in stochastic resonance}
As mentioned in the introduction, the constructive role played by noise in non-linear systems has been exploited to realise many engineering applications. The incorporation of ferroelectrics in semiconductor devices elicits an investigation of this phenomenon. We study the possibility of observing SR in a ferroelectric capacitor in this section, and provide numerical simulation results to motivate an application for weak periodic signal detection. The relation between SR and Kramers' problem is also provided towards the end of the section. \par
\begin{figure}[t!]
\centering
\includegraphics[scale=0.75]{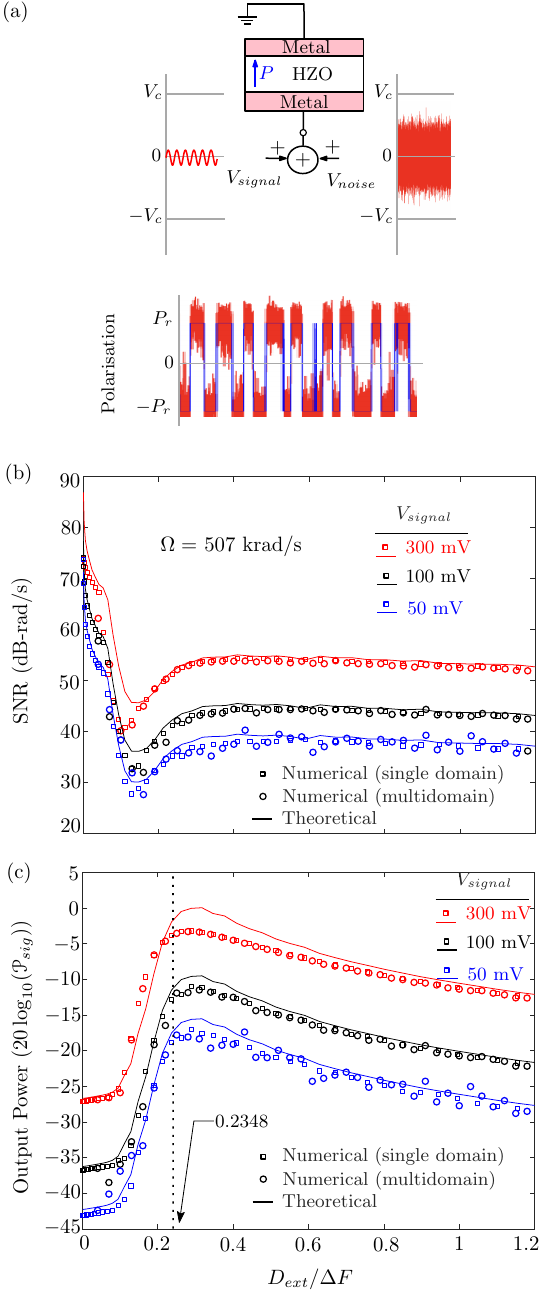}
\caption{(a) Schematic showing quasi-periodic switching of a ferroelectric due to a weak sinusoidal input (amplitude below $V_c$) and noise. (b,c) SNR and Output Power as a function of normalised noise intensity ($D_{ext}/\Delta F$).   The dotted line shows the approximate noise intensity value that yields maximum output power, determined from Kramers' theory (${\Omega}/{2\pi} \approx {r_K}/{2}$). }
\label{fig:snrfit}
\end{figure}

We are interested in determining the polarisation of the ferroelectric capacitor when a voltage comprising of noise and a weak periodic signal is applied, as shown in Figure \ref{fig:snrfit}(a).  The response of non-linear systems to stochastic inputs has been of significant interest to the mathematics and engineering communities. There are various approaches to this problem, involving different approximations \cite{gammaitoni_RMP_1998}. Linear Response Theory (LRT) \cite{luchinsky_ISCASPt1_1999},  treats the problem in two steps. First, the response of the non-linear system  to only the noise, is characterized. Then, the weak periodic input is treated as a perturbation, so that  the ensemble-averaged output is a scaled version of the periodic input. The scaling factor (a complex number, with magnitude and phase) depends on the  response of the non-linear system  to only the noise.  This is inferred from Bussgang's theorem~\cite{price_IRETransIT_1958}. Luchinsky et al. have exploited this idea and demonstrated SR in a circuit with a double-well landscape\cite{luchinsky_ISCASPt1_1999}. Along similar lines, we have investigated SR in ferroelectrics below. \par

Consider a  weak signal $V_{signal}(t) = V_0 \cos(\Omega t)$ given as an input to the ferroelectric as shown in Figure \ref{fig:snrfit}(a). A noise voltage, with root mean squared value $V_{noise}$, corresponding to noise strength $D_{ext}$  is added to the system (eq. (\ref{eq:D})).  In the framework of LRT, we first look at the system only with noise, but without the weak signal. Let $\chi(\omega; D_{ext})$ represent the electrical susceptibility (dimensionless) of the ferroelectric at a frequency $\omega$ with noise strength $D_{ext}$. For brevity, we drop the reference to $D_{ext}$ in $\chi(\omega; D_{ext})$ unless where required.  The susceptibility  $\chi(t)$  in   time-domain represents the impulse response of the system. Then, treating the weak applied electric field $E(t) = V(t)/t_F = E_0 \cos(\Omega t)$ (V/m) as a perturbation, we write the ensemble-averaged (denoted by $\langle \cdot \rangle$) output polarisation
\begin{equation}\label{eq:lrtout}
\langle P(t) \rangle  = \epsilon_0E_0|\chi(\Omega)|\cos(\Omega t + \phi)
\end{equation}
where
\begin{equation}\label{eq:phase}
\phi = -\tan^{-1}\left[ \frac{\Im\{\chi(\Omega)\}}{\Re\{\chi(\Omega)\}} \right].
\end{equation}
Here, $\epsilon_0$ is the permittivity of vacuum. $\Re\{(\cdot)\}$ and $\Im\{(\cdot)\}$ represent the real and imaginary components of a complex number.  
The objective of the analysis that follows is to obtain an expression for $\chi(\omega)$ using tools that are commonly employed in statistical physics and signal processing. \par

In accordance with Kubo's fluctuation-dissipation theorem \cite{kubo_repprogphy_1966}, we can relate the susceptibility $\chi(t)$ to the autocorrelation $R(t)$ of the polarisation of the unperturbed system (i.e  with noise but without sinusoidal input), 
\begin{equation}
\label{eq:flucdiskubotime}
\chi(t) = -\frac{1}{\epsilon_0D_{ext}}\frac{dR(t)}{dt}\Theta(t)
\end{equation}
where $\Theta(t)$ is the Heaviside function. Applying the Weiner-Khinchin theorem to eq. (\ref{eq:flucdiskubotime}), this equation may be expressed in the frequency domain, since power spectral density (PSD) is related to the Fourier transform of  the autocorrelation $R(t)$. \begin{equation}
\chi(\omega) = \frac{-1}{2 \pi\varepsilon_{0} D_{e x t}} \left[j \omega \frac{S_{N}^{0}(\omega) }{2}\ast  \left(\frac{1}{j \omega}+\pi \delta(\omega)\right)\right]
\end{equation}
where $S_N^0(\omega) \equiv S_N^0(\omega; D_{ext})$ is the one-sided PSD of the polarisation switching  in the unperturbed system, given by
\begin{equation}\label{eq:flucdiskubo}
S_N^0(\omega; D_{ext}) = \frac{4D_{ext}\epsilon_0}{\omega}\Im\{\chi(\omega; D_{ext})\}
\end{equation}
The two-sided power spectral density would be half of this term, since the power is spread over both positive and negative frequencies. Note that $\Im\{\chi(\omega)\}$ is associated with the  dissipation in the system. Inverting eq. (\ref{eq:flucdiskubo}), we can hence determine $\Im\{\chi(\omega)\}$ from the PSD, $S_N^0(\omega; D_{ext})$, of either the experimentally measured, or numerically simulated
polarisation of the unperturbed system. Given $\Im\{\chi(\omega)\}$, the real part can be found using the Kramers-Kronig relation \cite{warwick_PhysRev_1956,king_2009},
\begin{equation}\label{eq:krakro}
\Re\{\chi(\omega)\} = \frac{2}{\pi} \cdot \mathcal{P}\left\{\int_0^{\infty}\frac{\omega' \Im\{\chi(\omega')\}}{\omega'^2 - \omega^2}d\omega' \right\}
\end{equation}
where $\mathcal{P}\{\cdot\}$ denotes Cauchy's principal value integral, to account for the singularity at $\omega' = \omega$. Note that this relation between the real and imaginary parts is similar to the Hilbert transform in signal processing, which is a useful mathematical tool when dealing with causal signals \cite{johansson_thesis_1999}.  Thus, we can obtain the complete $\chi(\omega; D_{ext}) = \Re\{\chi(\omega; D_{ext})\} + j \Im\{\chi(\omega; D_{ext})\}$  from the response  of the unperturbed system. 

Going now to the perturbed system, the ensemble averaged output consists of a periodic signal component along with background noise. We wish to analytically characterize the relative strengths of the periodic signal component and the background noise. From eq. (\ref{eq:lrtout}), the power in the periodic signal component is given by $\mathscr{P}_{sig}(D_{ext}) = {\epsilon_0^2E_0^2|\chi(\Omega)|^2}/{2}$. We assume that the PSD of the noise background (around $\omega = \Omega$) in the  
output of the perturbed system, is identical to the PSD of the corresponding noise background in the unperturbed system. Hence, an analytical estimate of  Signal-to-Noise Ratio (SNR) at the output is 
\begin{equation}\label{eq:snrana}
\text{SNR}_{theory}(D_{ext}) = \frac{\mathscr{P}_{sig}(D_{ext})}{S_N^0(\Omega; D_{ext})}
\end{equation}
Note that since this is the ratio between a power and a PSD, it has units of dB-rad/s.  

We  also estimate the SNR numerically by taking the ratio of the signal power within a narrow band $[\Omega - \Delta \omega, \Omega + \Delta \omega]$,  to the noise background  at $\omega= \Omega$. This is the definition that is often used in the SR literature \cite{gammaitoni_RMP_1998}. 
\begin{equation}\label{eq:gamtni}
\text{SNR}_{num}(D_{ext}) = \lim_{\Delta \omega \to 0} \frac{\int_{\Omega-\Delta \omega}^{\Omega+\Delta \omega} S(\omega)d\omega - 2S_N(\Omega)\Delta \omega}{S_N(\Omega)},
\end{equation}
where $S(\omega) \equiv S(\omega; D_{ext})$ is the PSD of the perturbed system. We estimate the PSD of the noise background, $S_N(\Omega) \equiv S_N(\Omega; D_{ext}) $, by fitting several data points  of the noise floor around the signal peak with a straight line,  and evaluating the value at $\omega = \Omega$. The noise term in the numerator of eq. (\ref{eq:gamtni}) is negligible. 

Figure \ref{fig:snrfit} (b,c) presents our numerical and theoretical results for SR in a HZO based ferroelectric capacitor. The horizontal axis is depicted in terms of the dimensionless quantity $D_{ext}/\Delta F$. Following Ref. \cite{gammaitoni_RMP_1998}, $\Omega_0 \equiv 2\alpha/\rho$ defines a frequency scale in eq. (\ref{eq:langvn}). We choose $\Omega = 0.01 \times \Omega_0 \equiv 507$ krad/s (80.6 kHz) in our simulations. We use an ensemble of 400 copies of the PSDs to determine $\chi(\omega; D_{ext})$ required to estimate the theoretical and numerical SNR curves. Note the characteristic trend of SR in  Figure \ref{fig:snrfit}(b) - the SNR first drops with an increase in noise, but then rises with a further increase in noise.  The match between the theoretical and numerical results supports the use of LRT to understand  SR in HZO based ferroelectric capacitors.

\par
SR is very closely linked to Kramers' escape problem. The output power $\mathscr{P}_{sig}(D_{ext})$ is maximum \cite{gammaitoni_RMP_1998} when the frequency of the weak periodic signal ${\Omega}/{2\pi} \approx {r_K}/{2}$. Noting the relationship (eq. \ref{eq:D}) between $r_K$ and $D_{ext}$, the optimum noise strength that should be added depends on the frequency of the weka periodic signal that has to be detected. At this \emph{resonant} condition, the polarisation switching in the ferroelectric would be synchronised (with some phase shift, (eq. (\ref{eq:phase})) with the input signal, thereby giving the desired output with maximal power. For our choice of $\Omega = 0.01 \times \Omega_0$, we predict the optimum noise strength $D_{ext}/\Delta F = 0.2348$, which is consistent with our results presented in Figure \ref{fig:snrfit}(c).  

\subsection{Stochastic resonance in multidomain ferroelectrics}
As mentioned in the introduction, it is still an ongoing debate as to whether homogeneous switching occurs in experimentally grown HZO.  For this purpose, we also explore an inhomogeneous,  multidomain ferroelectric switching model in the context of SR.  Our results show good agreement with the results for the single domain case and with linear response theory. \par
\begin{figure}[t!]
\centering
\includegraphics[scale=0.7]{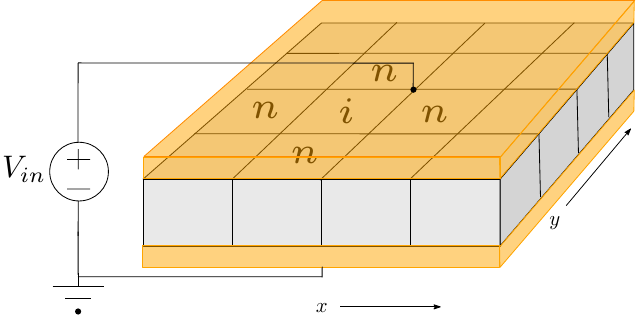}
\caption{Ferroelectric modelled as multiple domains which interact through the coupling constant $k$. }
\label{fig:mfm}
\end{figure}

The ferroelectric is modelled as a multidomain MFM (Metal-Ferroelectric-Metal) capacitor structure with $N (=400)$ domains. Every $i^{th}$ domain has a local spontaneous polarization $P_i$ and interacts with $n$ of its neighbouring domains, as shown in Fig.\ref{fig:mfm}.  
The Multi-domain Landau-Ginzburg model characterises the multidomain ferroelectric through the following equation: \cite{rollo_nano_2020,esseni_nano_2021}
\begin{equation} \label{eq:MDLK}
    \dot{P_i} = \frac{1}{\rho}\left[\frac{V_{in}}{t_F} - \alpha_{i} P_i - \beta_{i} P_i^3  - \frac{k}{dw}\sum_{n} (P_i - P_n)\right]
\end{equation}
where ${i}$ is the index of the domain (from 1 to $N$). $P_i$ and $P_n$ represent the polarisation of the $i^{th}$ domain and $n^{th}$ neighbouring domain respectively.  $k$, $d$ and $w$ denote the coefficient of gradient energy, domain width and domain wall width respectively.  $k$ is an intrinsic material parameter which usually lies between $10^{-9}$ - $10^{-11}$ $m^3/F$ \cite{esseni_nano_2021,cano_apl_2010}.  $\alpha_i$ and $\beta_i$  are used in order to account for the inhomogeneous nature of the ferroelectric. This is done by using the same $E_c$ and $P_r$ values for the single domain case and incorporating a deviation of 1\% from these values assuming a normal distribution.  \par 
For the same parameters as table \ref{tab:ferroparam} as well as the same voltage inputs and frequencies taken for the previous section,  the obtained SNR and output power plots match well with the data.  For each of 40 different noise intensities (once again, represented as a fraction of the barrier height $D/\Delta F$),  an ensemble average for 40 systems is considered.  The algorithm is similar to eq. \ref{eq:algo} with the extra coupling term in eq. \ref{eq:MDLK} as well as the usage of matrices instead of arrays in the euler algorithm.  Readers must also note that since there are $N=400$ domains,  simulations are more numerically intensive and therefore a smaller ensemble average is taken here as compared to the single domain case (i.e.,400).  These variations are evident from the results presented in Fig. \ref{fig:snrfit}. \par
It is to be noted that the primary distinction between the two cases under consideration is that the assumption of homogeneous switching provides a perspective through Kramers' escape problem.  Additionally, the incorporation of multiple ferroelectric domains in numerical simulations shows that the analysis for SR is not strictly limited to homogeneity.  This presents the possibility of realising applications which are driven by SR,  such as weak signal detection in communication systems.    

To summarize, we presented a comprehensive, quantitative analysis of stochastic resonance (SR) in ferroelectric materials.  Under the assumption of homogeneous switching according, the relation between SR and Kramers' escape problem in the context of a ferroelectric (Hf\textsubscript{0.5}Zr\textsubscript{0.5}O\textsubscript{2} or HZO) was investigated. We proposed an application that exploits the phenomenon of SR to potentially realise a weak periodic signal detector using this material system. The simulation results suggest that SR should occur and can be explained well using linear response theory.  Additionally, we validate that an optimum noise strength, related to Kramers rate $r_K$, can be chosen to  to maximize the output power.  To further demonstrate the feasibility of the single domain results, we presented and matched results for a multidomain case which is closer to realistic ferroelectric behaviour. 

\section*{acknowledgments}
The authors gratefully acknowledge the use of the CHANDRA High Performance Computing cluster at IIT Palakkad. AA thanks SERB (Science and Engineering Research Board, Government of India) for support through  MTR/2021/000823, CRG/2022/008128. AV thanks SERB Early Career Research Award (Grant No. ECR/2018/001076) for supporting the SURGE internship of MR. The authors also thank Dr. Lakshmi Narasihman Theagarajan and Dr. Debarati Chatterjee from IIT Palakkad for insightful discussions.



%
%

%


\bibliography{stochres}

\end{document}